\documentclass[12pt, preprint]{aastex}

\slugcomment{published in \emph{Geophysical Research Letters}, \textbf{38}: L24206 (2011) | doi: 10.1029/2011GL049806}

\shorttitle{Measurements of Martian Dust Devil Winds}
\shortauthors{Choi and Dundas}

\begin{document}

\title{Measurements of Martian Dust Devil Winds with HiRISE}

\author{David S. Choi}
\affil{ORAU/NASA Goddard Space Flight Center, Greenbelt, Maryland, USA.}
\email{david.s.choi@nasa.gov}

\and

\author{Colin M. Dundas}
\affil{United States Geological Survey, Astrogeology Science Center, Flagstaff, Arizona, USA.}

\begin{abstract}
We report wind measurements within Martian dust devils observed in plan view from the High Resolution Imaging Science Experiment (HiRISE) orbiting Mars. The central color swath of the HiRISE instrument has three separate charge-coupled devices (CCDs) and color filters that observe the surface in rapid cadence. Active features, such as dust devils, appear in motion when observed by this region of the instrument. Our image animations reveal clear circulatory motion within dust devils that is separate from their translational motion across the Martian surface. Both manual and automated tracking of dust devil clouds reveal tangential winds that approach 20-30 m s$^{-1}$ in some cases. These winds are sufficient to induce a $\sim$1\% decrease in atmospheric pressure within the dust devil core relative to ambient, facilitating dust lifting by reducing the threshold wind speed for particle elevation. Finally, radial velocity profiles constructed from our automated measurements test the Rankine vortex model for dust devil structure. Our profiles successfully reveal the solid body rotation component in the interior, but fail to conclusively illuminate the profile in the outer regions of the vortex. One profile provides evidence for a velocity decrease as a function of r$^{-1/2}$, instead of r$^{-1}$, suggestive of surface friction effects. However, other profiles do not support this observation, or do not contain enough measurements to produce meaningful insights. 
\end{abstract}

\pagebreak

%
%

\section{Introduction}

Measurements of the winds within Martian dust devils \citep{Balmerev06} are challenging to obtain. Some knowledge of their winds stems from meteorological instrumentation on the \emph{Viking} landers \citep{Hess77} as vortices either approached or passed directly over the lander. \citet{Ringrose03} analyzed the \emph{Viking} data and derived vortex core maximum wind speeds above 40 m s$^{-1}$ for certain dust devils. However, this analysis carries some uncertainty because it is dependent on predictions of the dust devil's distance from the instrumentation, as well as assumptions regarding its radial velocity profile. Other studies utilizing rover-based \citep{Metzger99, Greeley06, Greeley10} or spacecraft \citep{Stanzel08} observations are typically capable of reporting only the translational speed at which a dust devil travels across the Martian surface. Measurements of tangential winds \emph{within} a dust devil are rare, as \citet{Cantor06} report a single measurement of tangential winds just below 15 m s$^{-1}$ in a dust devil observed by the \emph{Viking} orbiter. 

Here we describe wind measurements within Martian dust devils using images taken with the HiRISE camera\citep{McEwen07} on board NASA's \emph{Mars Reconnaissance Orbiter}. Although the instrument's primary objective is the observation of geologic features rather than dynamic meteorological phenomena, we utilize both the instrument's method for building color images of the surface as well as serendipitous observations of active dust devils to quantify their motion.  

\section{Method}

The HiRISE camera utilizes a central color swath. This region features separate CCDs that image the surface through blue-green, red, and IR filters. During an observation, each CCD does not observe simultaneously; instead, a short time interval ($\sim$0.1 s) separates images taken through different filters. Thus, active features, such as dust devils or avalanches \citep{Russell08}, observed by the color swath appear in motion when animating the component images (see Supplementary Material). Tracking of dust devil cloud contrast features between frames yields horizontal wind measurements from orbit.

We identified 4 candidate dust devils exhibiting sufficient motion and cloud feature contrast. We tracked features manually by selecting a contrast feature in one image and identifying the same feature in a later image. We then calculated the offset distance in pixels between the selected points; this offset is the basis for the velocity measurement. Optimal manual measurements arise from dust clouds with moderate optical depth. Low optical depth clouds yield faint contrast features that are difficult to track, whereas high optical depth clouds can yield `'washed-out'' features that are similarly problematic.

Though recent advances in automated cloud feature tracking improve our measurements of atmospheric winds on the jovian planets \citep{Choi10}, this method does not consistently measure dust devil motion with success. Because these automated algorithms rely on optimizing correlations between cropped image portions (correlation windows), the background static Martian surface tends to dominate the comparison, frequently resulting in apparent velocity measurements near zero. However, we obtain some success in areas with more substantial dust clouds and relatively featureless background terrain.   

\section{Results}

\subsection{Wind vectors}

Table \ref{Table: dd_properties} lists the physical properties of the dust devils in our data set. Dust devils in our data set range from 30 to 250 m in diameter, and from 150 to 700 m in height. \citet{Greeley10} report that these diameters are typical for dust devils on the surface of Gusev crater, whereas \citet{Stanzel08} describe \emph{Mars Express} observations of dust devils similar in size to the largest dust devil in our data. Because of orbital constraints, all HiRISE images observe the surface at $\sim$3 PM local Mars time. In Gusev crater, dust devil activity on Mars tends to diminish by late afternoon from a peak around 1 PM \citep{Greeley10}. However, \emph{Mars Global Surveyor} observed considerable activity occurring in its 2 PM orbit \citep{Fisher05}, and \emph{Mars Express} notes a similar peak in activity between 2 and 3 PM local time \citep{Stanzel08}. 

\begin{figure}
\noindent\includegraphics[width=39pc, angle=90]{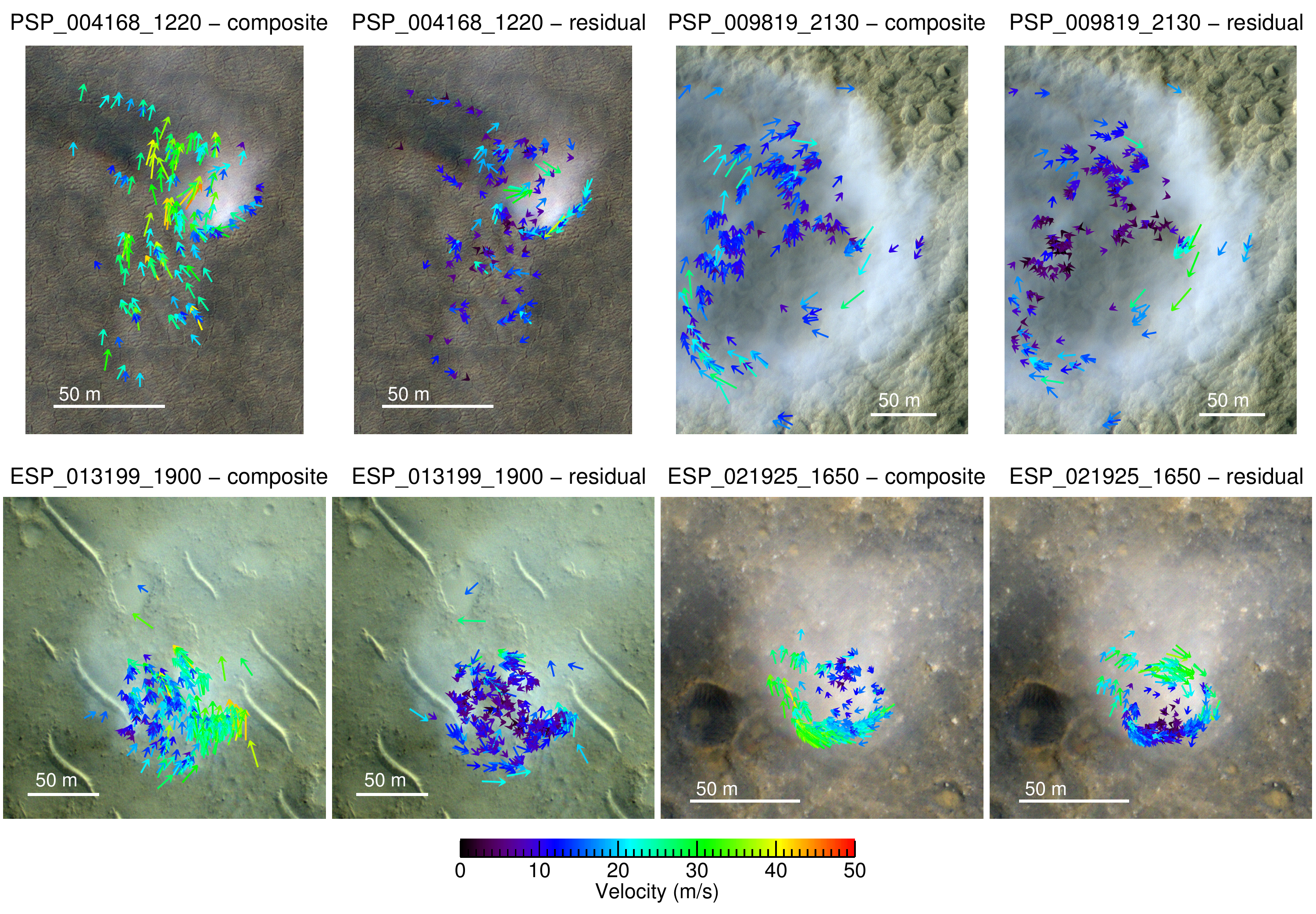}
\caption{Composite (raw) and residual (corrected for translational motion) wind measurements overlain on IRB (infrared, red, blue-green) color composite images of the dust devils analyzed in this work. The center of each vector corresponds to the location of the wind measurement. Both the color and length of the vector represent wind magnitude.
\label{Figure: vecmap}}
\end{figure}

Figure \ref{Figure: vecmap} shows two sets of wind vectors for each dust devil: manual measurements representing composite translational and tangential motion, and derived measurements of tangential motion only. All manual wind measurements shown in Figure \ref{Figure: vecmap} stem from all three possible image pairs for each dust devil. Separation of the wind vector into its translational and tangential components is an issue because the overall observation is short ($\sim$0.2 s), and tracking dust devils across the surface over timescales of minutes is not possible. Instead, we estimate translational velocity by calculating the overall average wind vector from the composite measurements. We subtract this estimate from the composite vector and exhibit the residual vectors in Figure \ref{Figure: vecmap} as the second set of wind vectors for each dust devil. Note that an accurate estimate using this method depends on an even spatial distribution of vectors from a symmetric dust devil in order for tangential motions to cancel out. Because some dust devils in our study better fit this criteria than others, it is important to consider that asymmetrical dust devils or irregularly distributed measurements will affect the velocity correction in these cases.

Overall, our composite wind measurements typically have a magnitude between 20 and 30 m s$^{-1}$, with a maximum near 45 m s$^{-1}$. Typically, the strongest winds occur along the outer edge of a dust devil, regardless of its diameter. When decomposing these measurements, we note some examples of strong translational winds. First, most measurements in PSP\_004168\_1220 are typically between 20 and 30 m s$^{-1}$ but likely possess a larger translational wind component because they originate from wispy dust clouds well outside of the central rotating column. In addition, composite vectors for the dust devil in ESP\_013199\_1900 reveal a gradation in wind magnitude from left to right and strong curl, signaling the effects of a considerably strong ambient wind on the inherent dust devil circulation. Our estimate of the translational wind speed is $\sim$20 m s$^{-1}$. These estimates are slightly above typical traverse speeds (15 m s$^{-1}$ and below) reported by \citet{Stanzel08}, though both \citet{Stanzel08} and \citet{Greeley10} report strong translational winds in a few instances, with some dust devils traveling at up to $\sim$50 m s$^{-1}$ across the Martian surface. 

\begin{figure}
\noindent\includegraphics[width=39pc]{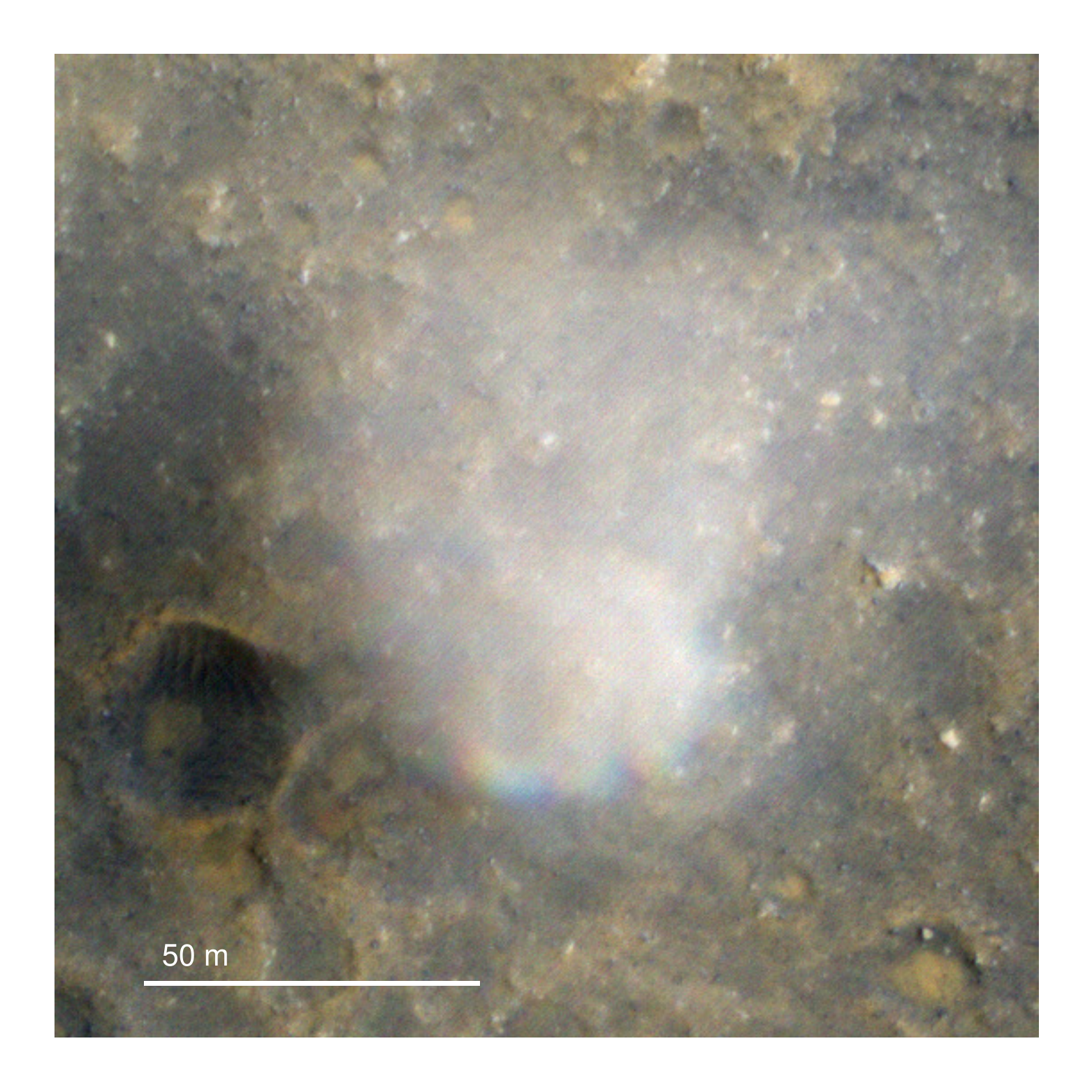}
\caption{IRB color composite of ESP\_021925\_1650. Dust clouds in motion produce rainbow fringes towards the bottom of the dust devil. 
\label{Figure: esp_021}}
\end{figure}

In other cases, the tangential component likely dominates the measurement. The dust devil in PSP\_009819\_2130 exhibits nearly closed circulation and a relatively low average wind magnitude ($\sim$9 m s$^{-1}$). This suggests weak ambient winds, though cropping of the dust devil and the spatial imbalance of measurements may yield an overestimate of the translational wind. Finally, the dust devil captured in ESP\_021925\_1650 (Figure \ref{Figure: esp_021}) displays substantial motion along its outer edge that produces rainbow ``fringe'' effects in the color composite image. Though imbalance in measurement location again generates overcorrected residual vectors, the quasi-closed circulation in the composite wind measurements (akin to what is seen in PSP\_009819\_2130) implies that ambient winds are relatively weak. This is in agreement with \citet{Greeley10}, who demonstrate that dust devil traverse speeds are $\ge$ 5 m s$^{-1}$ within Gusev crater.

\subsection{Radial Profiles}

\begin{figure}
\noindent\includegraphics[width=39pc]{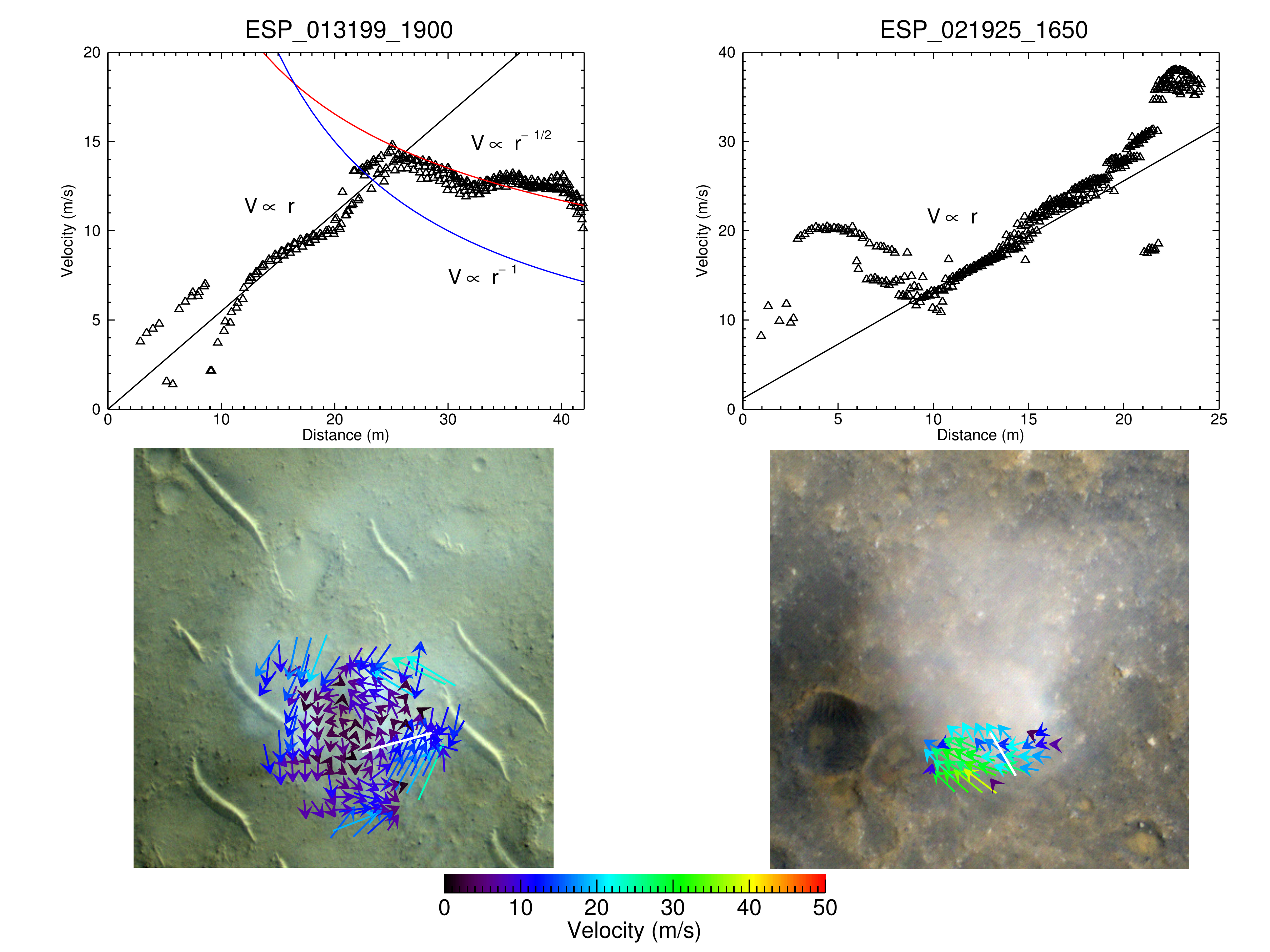}
\caption{Discrete radial profiles of wind measurements from ESP\_013199\_1900 (residual) and ESP\_021925\_1650 (non-residual), along with automated wind measurements of both dust devils. Wind measurements plotted in the profile stem from the white line shown in the wind vector maps. The maps show only a small percentage of wind measurements in the data set for clarity.
\label{Figure: radprofile}}
\end{figure}

Both ESP\_013199\_1900 and ESP\_021925\_1650 yielded successful measurements from our automated feature tracking software. The bottom row of Figure \ref{Figure: radprofile} shows maps displaying these automated results, though only a small percentage of the measurements is shown for clarity. Our software performs measurements at every possible pixel in an image except along edges, where the width of the excluded area along the edges depends on the correlation window size. The substantially larger number of measurements (typically 2 orders of magnitude above the number of manually tracked measurements) as a result allows construction of meaningful radial profiles. (However, the number of independent measurements is less because measurements located at adjacent pixels use a base correlation window that is nearly indistinguishable.)

\citet{Sinclair73} reported good agreement of terrestrial dust devil measurements with a Rankine vortex model. In a Rankine vortex, the core exhibits solid body rotation, with tangential velocity linearly increasing with radius. Outside of the radius with peak tangential winds, the velocity decreases as a function of r$^{-1}$. Our measurements confirm this region of solid body rotation, with radial profiles at several azimuths for the dust devils in ESP\_013199\_1900 and ESP\_021925\_1650 showing a linear increase in tangential winds. When defining the circulation center as the point where tangential velocities are closest to zero, the radius of maximum winds is $\sim$30 $\pm$ 5 m for both dust devils. This corresponds to the outer visible edge of the dust devil in ESP\_021925\_1650, but is slightly within the visible edge of the dust devil in ESP\_013199\_1900. Unfortunately, measurements outside of this area of solid body rotation are unsuccessful in most cases, and we cannot completely illuminate the velocity structure of the dust devil. 

One such profile, however, shown in Figure \ref{Figure: radprofile}, upper left, reveals a gradual r$^{-1/2}$ velocity decrease outside of the central vortex instead of the theoretical r$^{-1}$. Intriguingly, \citet{Tratt03} observed a similar profile of tangential winds within terrestrial dust devils. Both \citet{Tratt03} and \citet{Balmerev06} attribute the r$^{-1/2}$ profile with surface friction causing nonconservation of angular momentum, because the densest dust clouds tend to be located near the surface. We stress caveats regarding this r$^{-1/2}$ profile, as this particular profile is isolated to a narrow range of azimuths; other profiles do not confirm this velocity structure because of insufficient measurement coverage or lack of cloud tracers exterior to the region of solid body rotation. As laboratory studies \citep{Greeley03} confirm the Rankine velocity profile, such departures from the Rankine model are likely confined to the surface boundary layer and are perhaps indicative of small-scale turbulent motion or vortex development within the central core. 

\section{Error Analysis}
\label{Section: Errors}

Typical displacements of contrast features between image frames are 5--10 pixels, with some above 20 pixels. The most significant source of measurement uncertainty is our imperfect manual tracking technique. We estimate a 1--2 pixel measurement error accrued when selecting common points within contrast features. This results in a $\sim$2--5 m s$^{-1}$ uncertainty for each wind measurement, except for measurements from ESP\_013199\_1900, where lower spatial resolution increases the uncertainty to $\sim$5--10 m s$^{-1}$. Measurement uncertainty from the automated technique, however, is below 1 pixel.

Because there is a slight variation in the emission angle in each color component image originating from the spacecraft's motion in the time interval between frames, an important correction to the raw measurements is compensating for parallax motion of the dust clouds, as Martian dust devils can potentially reach heights of $\sim$8 km \citep{Fisher05}. In order to simplify the parallax correction, we utilize non map-projected images for analysis. In these images, the vertical axis of the image is parallel with the spacecraft ground track, with the direction of spacecraft motion towards the bottom of each image. This confines the induced apparent motion from parallax towards the top of each image. 

We estimated dust devil heights by measuring shadow lengths and accounting for illumination geometry. We cannot constrain the heights of the tracked dust clouds, so for parallax correction we assume that they are at the mid-height of the dust devil. (An exception is for the features tracked in PSP\_009819\_2130, where darker regions of the dust devil shadow imaged by HiRISE outside of the color swath likely correspond to the tracked dust clouds.) However, if the r$^{-1/2}$ radial profile in Figure \ref{Figure: radprofile} signifies that most dust clouds are closer to the surface, the true parallax effect may be rather small, rendering our overall velocity measurements as a conservative lower limit. Regardless, most offset corrections are below 1 pixel, corresponding to a velocity correction below $\sim$2--5 m s$^{-1}$. All results discussed in this paper and shown in Figures \ref{Figure: vecmap} and \ref{Figure: radprofile} properly account for parallax. 

One cautionary point is that a dust devil contrast feature may appear different in individual frames when observed with different CCD filters. However, contrast optimization and image rescaling yielded no significant difference in the measurements compared with the default processing pipeline. Another caveat is the time-delay integration (TDI) technique that HiRISE employs to improve signal-to-noise. The light collection time for a given pixel is as much as $\sim$11\% of the time interval between frames, though this time may vary for different filters within the same observation. However, there is no evidence of defects in the automated results from smearing, and selection bias prevents affected features from inclusion during manual tracking.

Finally, HiRISE is a line-scan instrument; the time interval between whole frames applies for lines at the same absolute ground position. Therefore, any motion of a dust devil feature parallel to the spacecraft ground track (and thus, perpendicular to the CCD lines) causes observation of said feature at a slightly shifted interval. However, the magnitude of this shift is two orders of magnitude below the time interval, and we ignore it for simplicity. In addition, HiRISE image processing applies a slight warping to images to register the color component images accurately and minimize spacecraft jitter. The background terrain in all of our image frames is well-registered, except for PSP\_009819\_2130, which retains residual jitter of $\sim$0.2--0.5 pixels along certain areas away from the dust devil. The jitter elevates the measurement uncertainty for that image, but its effect is minor relative to the dust devil motion after correcting for parallax. 

\section{Discussion}
\label{Section: Discussion}

Accurate assessments of winds within dust devils are important in order to constrain their efficiency at dust lifting, which carries implications into the role of dust devils in the Martian climate and maintaining the background atmospheric haze \citep{Newman02, Basu04}. Measurements indicate that the fine, entrained dust in the Martian atmosphere is $\sim$2 $\mu$m in diameter \citep{Tomasko99}. However, wind tunnel studies by \citet{Iversen82} demonstrate that boundary-layer winds necessary to move 2 $\mu$m dust via shear stress exceed those observed or predicted for the Martian surface by an order of magnitude. 

Thus, the precise mechanism by which dust devils raise dust into the atmosphere has inspired several studies. \citet{Greeley03} built a laboratory apparatus that simulated terrestrial and Martian dust devil vortices in order to study particle lifting threshold. They found that the decrease in pressure within the dust devil core facilitates the lift of particles (the $\Delta$P effect), reducing the threshold wind speed necessary to elevate particles compared to boundary-layer winds acting alone. \citet{Balme06} provide theoretical arguments that the $\Delta$P effect primarily operates via pressure gradients causing sudden, explosive release of subsurface air pockets beneath the dust, though their study could not conclusively determine if Martian soil would be able to degas and re-establish pressure equilibrium, preventing lifting from occurring.

Regardless, our measurements of dust devil velocity support a significant $\Delta$P, which could affect dust lifting into the Martian atmosphere. \citet{Greeley03} estimate that lifting of 2$\mu$m dust in atmospheric conditions similar to Mars require tangential wind speeds up to 20--30 m s$^{-1}$, but are likely lower on Mars due to its reduced gravity. All dust devils in our study exhibit areas with tangential winds above 10 m s$^{-1}$, though the prevalence of winds exceeding the threshold for dust lifting is variable. In addition, estimating the pressure perturbation of a dust devil, arguably a more important factor in the lifting of dust particles, is straightforward from velocity measurements. When assuming cyclostrophic balance and solid body rotation in the core, tangential winds of 10--20 m s$^{-1}$ yield a 0.25--1\% pressure decrease in the dust devil core relative to ambient pressure. This agrees with several studies of both terrestrial \citep{Sinclair73}, and Martian dust devils \citep{Metzger99, Ringrose03}. 

Our initial search of archival HiRISE data for dust devils yielded four other dust devils that were unsuitable for feature tracking because of poor contrast. However, future comprehensive searches for previously observed but unknown dust devils may supplement our knowledge of dust devil velocities. In addition, enhancements to our automated technique may increase the number of effective wind measurements on the current data set. Subsequent HiRISE observations of regions with known dust devil activity can also further expand our data set, though successful observations may be rare. Given the infrequency of dedicated meteorological instruments on surface rovers, however, an orbital imaging strategy for the monitoring and measurement of Martian dust devils may be an optimal plan going forward.

\acknowledgments
We thank two anonymous referees for constructive comments that improved this manuscript. This work was supported by the Mars Reconnaissance Orbiter Project and NASA Grant \#NNX09AD98G. We thank the HiRISE team for planning and acquiring the images used in this work. We also thank Laszlo Kestay, Moses Milazzo, Paul Geissler, Randy Kirk, Ingrid Daubar Spitale, Adam Showman, Emily Rauscher, and Matt Balme for helpful discussions.

\begin{table*}
\begin{tabular}{lrrcrr}
Image ID & Latitude & Longitude (east) & L$_s$ & Diameter (m) & Height (m) \\
PSP\_004168\_1220 & -57.9$^{\circ}$ & 65.5$^{\circ}$ & 259.1$^{\circ}$ & $\sim$25 & $\sim$150\\
PSP\_009819\_2130 & 32.8$^{\circ}$ & 199.5$^{\circ}$ & 120.0$^{\circ}$ & $\sim$250 & $\sim$650\\
ESP\_013199\_1900 & 9.8$^{\circ}$ & 83.1$^{\circ}$ & 269.7$^{\circ}$ & $\sim$100 & $\sim$400\\
ESP\_021925\_1650 & -14.6$^{\circ}$ & 175.5$^{\circ}$ & 265.3$^{\circ}$ & $\sim$50 & $\sim$150\\
\end{tabular}
\caption{Physical properties of the Martian dust devils examined in this study. Note that the height listed for PSP\_009819\_2130 is for the entire dust devil, whereas we estimate the tracked dust clouds to be at a height $\leq$ 250 m, corresponding to the densest areas of its shadow. L$_s$ is the Mars-Sun angle (Martian solar longitude), with L$_s$ = 0$^{\circ}$ defined as Northern Hemisphere vernal equinox.
\label{Table: dd_properties}}
\end{table*}



%
%
%
%
%
%



\begin{thebibliography}{20}
\providecommand{\natexlab}[1]{#1}
\expandafter\ifx\csname urlstyle\endcsname\relax
  \providecommand{\doi}[1]{doi:\discretionary{}{}{}#1}\else
  \providecommand{\doi}{doi:\discretionary{}{}{}\begingroup
  \urlstyle{rm}\Url}\fi

\bibitem[{\textit{{Balme} and {Greeley}}(2006)}]{Balmerev06}
{Balme}, M., and R.~{Greeley} (2006), {Dust devils on Earth and Mars},
  \textit{Reviews of Geophysics}, \textit{44}, RG3003,
  \doi{10.1029/2005RG000188}.

\bibitem[{\textit{{Balme} and {Hagermann}}(2006)}]{Balme06}
{Balme}, M., and A.~{Hagermann} (2006), {Particle lifting at the soil-air
  interface by atmospheric pressure excursions in dust devils}, \textit{\grl},
  \textit{33}, L19S01, \doi{10.1029/2006GL026819}.

\bibitem[{\textit{{Basu} et~al.}(2004)\textit{{Basu}, {Richardson}, and
  {Wilson}}}]{Basu04}
{Basu}, S., M.~I. {Richardson}, and R.~J. {Wilson} (2004), {Simulation of the
  Martian dust cycle with the GFDL Mars GCM}, \textit{Journal of Geophysical
  Research (Planets)}, \textit{109}, E11,006, \doi{10.1029/2004JE002243}.

\bibitem[{\textit{{Cantor} et~al.}(2006)\textit{{Cantor}, {Kanak}, and
  {Edgett}}}]{Cantor06}
{Cantor}, B.~A., K.~M. {Kanak}, and K.~S. {Edgett} (2006), {Mars Orbiter Camera
  observations of Martian dust devils and their tracks (September 1997 to
  January 2006) and evaluation of theoretical vortex models}, \textit{Journal
  of Geophysical Research (Planets)}, \textit{111}, E12,002,
  \doi{10.1029/2006JE002700}.

\bibitem[{\textit{{Choi} et~al.}(2010)\textit{{Choi}, {Showman}, and
  {Vasavada}}}]{Choi10}
{Choi}, D.~S., A.~P. {Showman}, and A.~R. {Vasavada} (2010), {The evolving flow
  of Jupiter's White Ovals and adjacent cyclones}, \textit{Icarus},
  \textit{207}, 359--372, \doi{10.1016/j.icarus.2009.10.013}.

\bibitem[{\textit{{Fisher} et~al.}(2005)\textit{{Fisher}, {Richardson},
  {Newman}, {Szwast}, {Graf}, {Basu}, {Ewald}, {Toigo}, and
  {Wilson}}}]{Fisher05}
{Fisher}, J.~A., M.~I. {Richardson}, C.~E. {Newman}, M.~A. {Szwast}, C.~{Graf},
  S.~{Basu}, S.~P. {Ewald}, A.~D. {Toigo}, and R.~J. {Wilson} (2005), {A survey
  of Martian dust devil activity using Mars Global Surveyor Mars Orbiter Camera
  images}, \textit{Journal of Geophysical Research (Planets)}, \textit{110},
  E03,004, \doi{10.1029/2003JE002165}.

\bibitem[{\textit{{Greeley} et~al.}(2003)\textit{{Greeley}, {Balme}, {Iversen},
  {Metzger}, {Mickelson}, {Phoreman}, and {White}}}]{Greeley03}
{Greeley}, R., M.~R. {Balme}, J.~D. {Iversen}, S.~{Metzger}, R.~{Mickelson},
  J.~{Phoreman}, and B.~{White} (2003), {Martian dust devils: Laboratory
  simulations of particle threshold}, \textit{Journal of Geophysical Research
  (Planets)}, \textit{108}, 5041, \doi{10.1029/2002JE001987}.

\bibitem[{\textit{{Greeley} et~al.}(2006)}]{Greeley06}
{Greeley}, R., et~al. (2006), {Active dust devils in Gusev crater, Mars:
  Observations from the Mars Exploration Rover Spirit}, \textit{Journal of
  Geophysical Research (Planets)}, \textit{111}, E12S09,
  \doi{10.1029/2006JE002743}.

\bibitem[{\textit{{Greeley} et~al.}(2010)\textit{{Greeley}, {Waller}, {Cabrol},
  {Landis}, {Lemmon}, {Neakrase}, {Pendleton Hoffer}, {Thompson}, and
  {Whelley}}}]{Greeley10}
{Greeley}, R., D.~A. {Waller}, N.~A. {Cabrol}, G.~A. {Landis}, M.~T. {Lemmon},
  L.~D.~V. {Neakrase}, M.~{Pendleton Hoffer}, S.~D. {Thompson}, and P.~L.
  {Whelley} (2010), {Gusev Crater, Mars: Observations of three dust devil
  seasons}, \textit{Journal of Geophysical Research (Planets)}, \textit{115},
  E00F02, \doi{10.1029/2010JE003608}.

\bibitem[{\textit{{Hess} et~al.}(1977)\textit{{Hess}, {Henry}, {Leovy},
  {Tillman}, and {Ryan}}}]{Hess77}
{Hess}, S.~L., R.~M. {Henry}, C.~B. {Leovy}, J.~E. {Tillman}, and J.~A. {Ryan}
  (1977), {Meteorological results from the surface of Mars - Viking 1 and 2},
  \textit{\jgr}, \textit{82}, 4559--4574, \doi{10.1029/JS082i028p04559}.

\bibitem[{\textit{{Iversen} and {White}}(1982)}]{Iversen82}
{Iversen}, J.~D., and B.~R. {White} (1982), {Saltation threshold on Earth, Mars
  and Venus}, \textit{Sedimentology}, \textit{29}, 111--119,
  \doi{10.1111/j.1365-3091.1982.tb01713.x}.

\bibitem[{\textit{{McEwen} et~al.}(2007)}]{McEwen07}
{McEwen}, A.~S., et~al. (2007), {Mars Reconnaissance Orbiter's High Resolution
  Imaging Science Experiment (HiRISE)}, \textit{Journal of Geophysical Research
  (Planets)}, \textit{112}, E05S02, \doi{10.1029/2005JE002605}.

\bibitem[{\textit{{Metzger} et~al.}(1999)\textit{{Metzger}, {Carr}, {Johnson},
  {Parker}, and {Lemmon}}}]{Metzger99}
{Metzger}, S.~M., J.~R. {Carr}, J.~R. {Johnson}, T.~J. {Parker}, and M.~T.
  {Lemmon} (1999), {Dust devil vortices seen by the Mars Pathfinder camera},
  \textit{\grl}, \textit{26}, 2781--2784, \doi{10.1029/1999GL008341}.

\bibitem[{\textit{{Newman} et~al.}(2002)\textit{{Newman}, {Lewis}, {Read}, and
  {Forget}}}]{Newman02}
{Newman}, C.~E., S.~R. {Lewis}, P.~L. {Read}, and F.~{Forget} (2002), {Modeling
  the Martian dust cycle, 1. Representations of dust transport processes},
  \textit{Journal of Geophysical Research (Planets)}, \textit{107}, 5123,
  \doi{10.1029/2002JE001910}.

\bibitem[{\textit{{Ringrose} et~al.}(2003)\textit{{Ringrose}, {Towner}, and
  {Zarnecki}}}]{Ringrose03}
{Ringrose}, T.~J., M.~C. {Towner}, and J.~C. {Zarnecki} (2003), {Convective
  vortices on Mars: a reanalysis of Viking Lander 2 meteorological data, sols
  1-60}, \textit{Icarus}, \textit{163}, 78--87,
  \doi{10.1016/S0019-1035(03)00073-3}.

\bibitem[{\textit{{Russell} et~al.}(2008)}]{Russell08}
{Russell}, P., et~al. (2008), {Seasonally active frost-dust avalanches on a
  north polar scarp of Mars captured by HiRISE}, \textit{\grl}, \textit{35},
  L23,204, \doi{10.1029/2008GL035790}.

\bibitem[{\textit{{Sinclair}}(1973)}]{Sinclair73}
{Sinclair}, P.~C. (1973), {The Lower Structure of Dust Devils.},
  \textit{Journal of Atmospheric Sciences}, \textit{30}, 1599--1619,
  \doi{10.1175/1520-0469(1973)030$<$1599:TLSODD$>$2.0.CO;2}.

\bibitem[{\textit{{Stanzel} et~al.}(2008)\textit{{Stanzel}, {P{\"a}tzold},
  {Williams}, {Whelley}, {Greeley}, {Neukum}, and {The HRSC Co-Investigator
  Team}}}]{Stanzel08}
{Stanzel}, C., M.~{P{\"a}tzold}, D.~A. {Williams}, P.~L. {Whelley},
  R.~{Greeley}, G.~{Neukum}, and {The HRSC Co-Investigator Team} (2008), {Dust
  devil speeds, directions of motion and general characteristics observed by
  the Mars Express High Resolution Stereo Camera}, \textit{Icarus},
  \textit{197}, 39--51, \doi{10.1016/j.icarus.2008.04.017}.

\bibitem[{\textit{{Tomasko} et~al.}(1999)\textit{{Tomasko}, {Doose}, {Lemmon},
  {Smith}, and {Wegryn}}}]{Tomasko99}
{Tomasko}, M.~G., L.~R. {Doose}, M.~{Lemmon}, P.~H. {Smith}, and E.~{Wegryn}
  (1999), {Properties of dust in the Martian atmosphere from the Imager on Mars
  Pathfinder}, \textit{\jgr}, \textit{104}, 8987--9008,
  \doi{10.1029/1998JE900016}.

\bibitem[{\textit{{Tratt} et~al.}(2003)\textit{{Tratt}, {Hecht}, {Catling},
  {Samulon}, and {Smith}}}]{Tratt03}
{Tratt}, D.~M., M.~H. {Hecht}, D.~C. {Catling}, E.~C. {Samulon}, and P.~H.
  {Smith} (2003), {In situ measurement of dust devil dynamics: Toward a
  strategy for Mars}, \textit{Journal of Geophysical Research (Planets)},
  \textit{108}, 5116, \doi{10.1029/2003JE002161}.

\end{thebibliography}
\end{document}